# All-nanophotonic NEMS biosensor on a chip

Dmitry Yu. Fedyanin* and Yury V. Stebunov

Laboratory of Nanooptics and Plasmonics, Moscow Institute of Physics and Technology,

Dolgoprudny, 141707 Russian Federation

* dmitry.fedyanin@phystech.edu

**Integrated chemical and biological sensors give advantages in cost, size and weight reduction and open new prospects for parallel monitoring and analysis. Biosensors based on nanoelectromechanical systems (NEMS) are the most attractive candidates for the integrated platform. However, actuation and transduction techniques (e.g. electrostatic, magnetomotive, thermal or piezoelectric) limit their operation to laboratory conditions. All-optical approach gives the possibility to overcome this problem, nevertheless, the existing schemes are either fundamentally macroscopic or excessively complicated and expensive in mass production. Here we propose a novel scheme of extremely compact NEMS biosensor monolithically integrated on a chip with all-nanophotonic transduction and actuation. It consists of the photonic waveguide and the nanobeam cantilever placed above the waveguide, both fabricated in the same CMOS-compatible process. Being in the near field of the strongly confined photonic mode, cantilever is efficiently actuated and its response is directly read out using the same optical waveguide, which results in a very high sensitivity and capability of single-molecule detection even in atmosphere.**

## Introduction

Advances in fabrication technologies make it possible to manufacture nanoscale chemical and biological sensors and integrate them in a wide range of different devices. This does not only decrease the size but also improves the resolution, reduces the price and opens the way to parallel monitoring and analysis on a chip. Nanoelectromechanical systems (NEMS) are considered to be the most attractive platform. NEMS biosensors are essentially compact, guarantee extremely high sensitivity with a detection down to the single molecule level and demonstrate a very broad range of possible applications from mass spectrometry to medical diagnostics [1-3]. Specific surfaces deposited on NEMS can adsorb detecting agents in vacuum, gaseous and fluid environments, which changes the oscillating mass and shifts the resonant frequency of mechanical oscillations. However, practical implementation of NEMS sensors is hindered by actuation and transduction schemes. Mechanical oscillations can be excited and detected using electrostatic [4, 5], magnetomotive [6], thermal [7-9], piezoresistive [10] and piezoelectric [11] techniques, but all these methods have significant disadvantages. For the magnetomotive actuation, a very high magnetic field is needed, which increases the device size and its price and makes it difficult to operate. Piezoelectric actuation at high resonant frequencies requires complex multilayer nanostructures with a thin piezoelectric layer. This impairs mechanical characteristics of the cantilever and affects chemical properties of its surface. Electrostatic schemes are rather inefficient at high resonant frequencies and small dimensions of NEMS. Thermal actuation and piezoresistive transduction induce heating, which significantly changes the kinetics of biomolecular reactions and consequently decreases the accuracy of measurements. In addition, application of the above-mentioned techniques in liquid and high pressure gas environments is seriously restrained by high leakage currents. All-optical approach can eliminate these drawbacks and simplify the design of NEMS biosensors.

Optical actuation and detection schemes based on a laser beam have been used for years [12, 13], but, being fundamentally macroscopic, they can hardly be a part of an integrated biosensor platform. Optical fibers confine light on the optical-wavelength scale and make the devices microscopic [14, 15]. This drastically increases sensitivity and accuracy of measurements. In spite of these advantages, such biosensors are quite bulky and expensive in mass production, which is dictated by the necessity to precisely position and align the fiber at a submicrometer distance from the NEMS. To develop a truly integrated NEMS biosensor, optical readout and actuation schemes should be fabricated directly on a chip, which can be achieved with planar waveguide technologies. Efficient readout of nanomechanical motion could be realized using coupling between guided and radiation modes of a suspended waveguide [16], coupling between a waveguide and an optical resonator [17, 18], end-to-end coupling of waveguides [19, 20], and side-coupling between a cantilever and a waveguide [21, 22]. Such configurations provide high displacement sensitivity and can be fabricated on a chip. However, their implementation is significantly restricted by the excessive complexity of fabrication processes.

Here, we present a novel type of strongly integrated on-chip NEMS biosensors with all-nanophotonic transduction and actuation. The proposed scheme shows extremely high sensitivity and efficiency and is capable of single-molecule detection even in atmosphere. For transduction of the nanobeam cantilever and its actuation we use the same nanophotonic waveguide, while the position of the cantilever above the waveguide can be controlled with sub-nanometer accuracy in the fabrication process using wet chemical etching. This ensures an exceedingly compact size, low cost fabrication and high reproducibility of the proposed integrated biosensor making it an ideal candidate for parallel monitoring and mass production.

## Results

**All-nanophotonic NEMS biosensor design and operating principle**

Fig. 1 shows the proposed configuration of the NEMS biosensor with all-nanophotonic transduction and actuation. It consists of only two basic elements: a nanophotonic waveguide and a free-oscillating cantilever placed above the waveguide a distance of a few hundred nanometers and aligned approximately perpendicular to the waveguide. In this case, the cantilever is in the waveguide near field, which induces mutual interaction between guided optical modes and the cantilever. First of all, the cantilever is a fairly large scatter for optical modes of the waveguide and, as the distance between the cantilever and the waveguide decreases, more radiation is scattered to free space and, consequently, less power is transmitted through the waveguide section with the cantilever. Accordingly, the oscillating cantilever modulates the intensity of the probe optical signal at a frequency of mechanical oscillations. Detecting molecules adsorbed on the surface of the cantilever change the resonant frequency of mechanical oscillations (dynamic regime) or create a surface tension resulting in cantilever bending (static regime). Both the magnitude of the transmitted optical signal and the frequency of modulation can be precisely measured with a photodetector and report the mass of adsorbed molecules and the adsorption rate, which forms an exceptionally efficient and accurate transduction scheme. Secondly, the electric field amplitude of the guided optical mode rapidly decreases with the distance from the waveguide core and the cantilever experiences a ponderomotive force, which is proportional to the gradient of the squared electric field. Thanks to the extremely high mode localization in nanophotonic waveguides, even a weak pump signal provide a few nanometers displacement of the cantilever. Being excited at a wavelength different from that of the probe signal and modulated at the resonant frequency of mechanical oscillations, such a low power pump signal actuates the NEMS without any effect on the transduction scheme. In contrast to electric and magnetic actuation techniques, the photonic approach does not introduce inherent limitations in the modulation bandwidth except that the electro-optic modulator is typically limited by a frequency of the order of 1 GHz, which is much higher than the resonant frequency of mechanical oscillations.

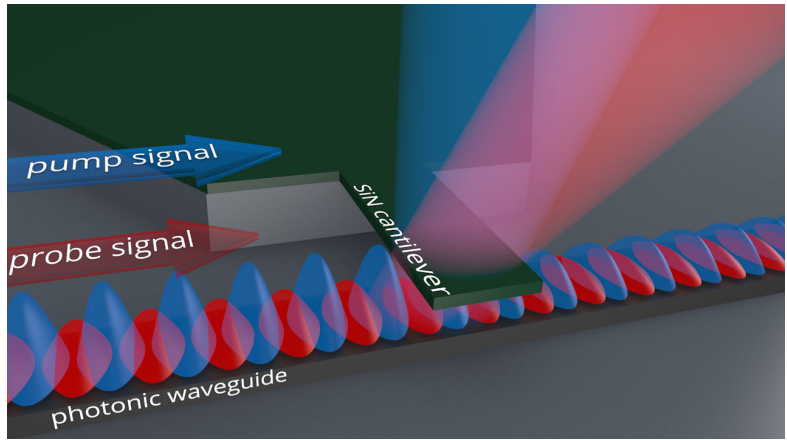

**Figure 1**. **Schematic view of the highly integrated all-nanophotonic NEMS biosensor.** Nanobeam cantilever is suspended above the photonic waveguide at a distance of a few hundred nanometers and is in the near-field of the optical mode of the waveguide. Pump optical signal excited at a light wavelength $\lambda_1$ and sinusoidally modulated at a frequency $f_M$ actuates the cantilever. At the same time, the power of the continuous wave probe signal excited at a light wavelength $\lambda_2$ and propagating along the same waveguide is controlled by the vibrating nanobeam, which gives a possibility to gauge the amplitude of mechanical oscillations. By doing the scan of the modulation frequency $f_M$, one can measure the decrease in the resonant frequency of the cantilever and determine the mass of adsorbed molecules.

Mechanical cantilever can be fabricated from different materials and have different shapes, but for on-chip integration, compatibility with conventional silicon CMOS fabrication processes and planar technologies is essential. This makes the nanobeam one of the best candidates, since it is very small in size, allows operation at high oscillation frequencies and is easily fabricated. Silicon seems to be the best material for the cantilever in photonic schemes owing to the high refractive index, but gas and biosensors typically operate in aggressive environments and chemical reactions (such as oxidation) on the silicon surface are practically unavoidable. They alter elasticity properties of the cantilever and change the oscillating mass along with the gas or biological molecules adsorbed on the cantilever, which significantly deteriorates the accuracy of measurements. In this regard, silicon nitride (SiN) is much preferred thanks to the high chemical stability and roughly the same mechanical properties as that of silicon. Small SiN beam with dimensions of $l \times w \times t$ clamped at one end and free at the other exhibits the lowest mechanical resonance at the frequency [23, 24]

$$f_M = \frac{3.516}{4\pi} \frac{t}{l^2} \sqrt{\frac{\Xi}{3\rho\left(1+\frac{4m}{\rho l w t}\right)}} \approx f_{M0} - R_f m, \qquad (1)$$

where $\Xi = 290$ GPa and $\rho = 3$ g/cm$^3$ [25] are the Young's modulus and mass density of the SiN beam, respectively, $f_{M0}$ is the resonant frequency of the cantilever, $m$ is the mass of molecules adsorbed at the end of the cantilever and $R_f$ is the mass responsivity. For the SiN beam dimensions of 5 μm×1 μm×90 nm, the quality factor of the resonance is about $Q_M = 3500$ [12], $f_{M0} = 5.72$ MHz and $R_f = 2f_{M0}/(\rho l w t) = 8.5 \times 10^{18}$ Hz/g.

Mode properties of the photonic waveguide are not as sensitive to ultrathin oxide layers on the silicon surface as that of mechanical ones of the cantilever, and the nanophotonic waveguide can be realized on a standard silicon-on-insulator (SOI) platform. Thus, 200 nm thick silicon strip on a silicon dioxide (SiO$_2$) layer confines light on the nanoscale and the penetration depth of the only guided TE$_0$ mode into the air is $\zeta_{top} = 96$ nm at $\lambda = 1.55$ μm, which is small enough for efficient transduction and actuation.

**Photonic transduction of the nanocantilever**

Optical mode propagating in the SOI waveguide passes through the section, where the SiN beam is located above the waveguide, and is partially reflected back, but mostly scattered into free space. When the distance $h$ between the nanobeam and the waveguide is much larger than the penetration depth $\zeta_{top}$ of the electromagnetic field of the optical mode into air or vacuum, the transmission coefficient $T$ is nearly equal to unity and there is no practical interaction between the cantilever and the optical mode. As the distance $h$ decreases, the SiN nanobeam changes significantly the refractive index profile of the waveguide and induce scattering. This can be treated as coupling between guided and radiation modes of the unperturbed waveguide [26-28] and can be expressed as

$$1-T = \int_0^{\omega/c} \left\{ \left|\int_0^w K_S(\beta)\exp[i(\beta_g-\beta)z]dz\right|^2 + \left|\int_0^w K_{AS}(\beta)\exp[i(\beta_g-\beta)z]dz\right|^2 \right.$$
$$\left. + \left|\int_0^w K_S(\beta)\exp[i(\beta_g+\beta)z]dz\right|^2 + \left|\int_0^w K_{AS}(\beta)\exp[i(\beta_g+\beta)z]dz\right|^2 \right\} \frac{\beta}{\sqrt{\frac{\omega^2}{c^2}-\beta^2}} d\beta \qquad (2)$$

where $\beta_g$ is the wavenumber of the guided mode and $K_S(\beta)$ and $K_{AS}(\beta)$ are the coupling coefficient between the guided mode and symmetric and anti-symmetric radiation modes, respectively:

$$K_S \approx \frac{\omega}{4\pi} \int_h^{h+t} (n_{SiN}^2 - 1) E_{gy}(x) E_{Sy}^*(\beta, x) dx \qquad (3)$$

$$K_{AS} \approx \frac{\omega}{4\pi} \int_h^{h+t} (n_{SiN}^2 - 1) E_{gy}(x) E_{ASy}^*(\beta, x) dx \qquad (4)$$

In the above expressions, $n_{SiN} = 1.98$ [29] is the refractive index of SiN, $h$ is the separation distance between the cantilever and the waveguide surface, $t$ is the thickness of the cantilever beam, $E_{gy}(x)$, $E_{Sy}(\beta,x)$ and $E_{ASy}(\beta,x)$ are the normalized [27] transverse electric field amplitudes of the guided mode, symmetric radiation mode with the wavenumber $\beta$ and anti-symmetric radiation mode with the wavenumber $\beta$, respectively. Equations (2) and (3) clearly show that the scattered power is approximately proportional to the square of electric-field amplitude at the position of the cantilever $|E_{gy}(h)|^2$ and decreases with the distance $h$ as $\exp(-2x/\zeta_{top})$. Accordingly, the oscillating beam modulates the total transmission with a modulation depth of $-2X_0\, dT/dx$, where $X_0$ is the amplitude of mechanical oscillations.

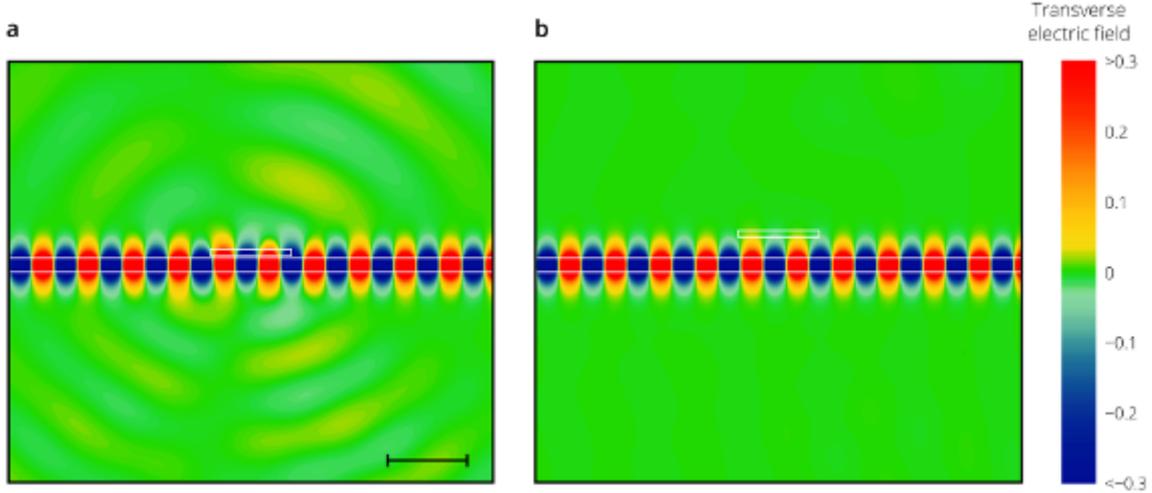

**Figure 2. Field distribution for the fundamental TE mode of the SOI waveguide passing through the section with the cantilever.** Transverse electric field distribution of the fundamental photonic mode guided by the 200 nm thick SOI waveguide and transmitted through the section with the SiN cantilever placed above the waveguide for two positions of the cantilever: 30 nm (panel a) and 300 nm (panel b) above the silicon waveguide. The scale bar is 1 μm and the light wavelength is 1.55 μm. At a separation distance of 30 nm between the waveguide and the cantilever, probe signal loses 0.35% of its power. As the distance between the cantilever beam and the waveguide increases, less power is scattered out and reflected back and, at a distance of 300 nm, the optical loss does not exceed 0.001%.

Transduction of the cantilever is accurately simulated using the 2D finite element method in COMSOL Multiphysics. The simulation domain is 25 μm in length, 14 μm in height and is surrounded by the perfectly matching layer. The mesh size is less than 20 nm near the waveguide and cantilever and is about 80 nm near the top and bottom boundaries of the simulation domain. Fig. 2 shows the distribution of the transverse electric field for two different positions of the cantilever. As the distance $h$ decreases, more energy of the guided optical mode is scattered by the cantilever both forward and backward (Fig. 2), which decreases the total transmission. The dependence of the transmitted power and the linear displacement detection responsivity $R_X = -\,dT/dx$ on the cantilever position is presented in Fig. 3a. The simulation results are in a good agreement with the prediction of the coupled mode theory. Despite that only $1 - T = 0.25\%$ of the mode power is lost at $h = 50$ nm, the responsivity $R_X = -0.04$ μm$^{-1}$ is higher than that of the fiber based systems [15], which creates the backbone for the realization of extremely compact and efficient gas and biosensors integrated on a chip.

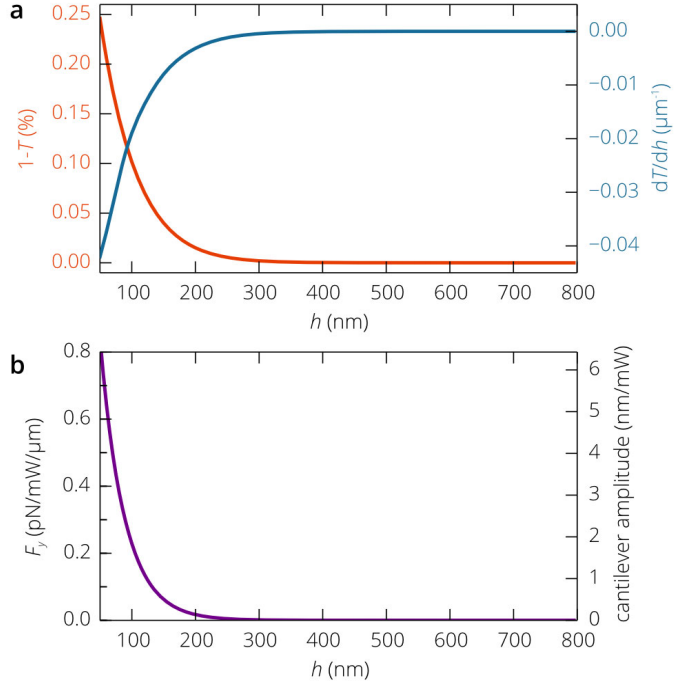

**Figure 3. Transduction and actuation of the SiN cantilever with the SOI waveguide.** (a) Simulated transmittance $T$ of the probe optical signal through the waveguide section with the cantilever (see Fig. 2) and its derivative $dT/dh$ as a function of the cantilever position. (b) Optical force generated on the cantilever by the continuous wave pump signal per unit power of the pump signal per unit length of the cantilever beam at a light wavelength of 1.31 μm and the amplitude of mechanical oscillations induced by the fully modulated pump optical signal propagating along the 1 μm wide silicon waveguide.

## Optomechanical actuation of the nanocantilever

The guided optical mode of the SOI waveguide is highly confined to the silicon core and its electric field decays exponentially with the distance from the silicon surface. The penetration depth into air is less than 100 nm at telecom wavelengths, which creates an extremely strong field gradient in the near field of the waveguide. Electric field polarizes the nanobeam and the positive and negative charges experience noticeably different forces due to the field gradient. Eventually, this results in a sufficiently large ponderomotive force acting on the cantilever, which can be exploited for actuation.

Net optical force $F$ generated on the cantilever includes the scattering force and the dominant gradient force and can be calculated by integrating the Maxwell's stress tensor $\mathbf{T}$ around the surface $S$ encapsulating the nanobeam [30]:

$$<\mathbf{F}> = \int_S \mathbf{n} <\mathbf{T}> ds \qquad (5)$$

Here $\mathbf{n}$ is the unit normal vector to the surface and the notation $<...>$ represents averaging over the period $c/\lambda$ of optical oscillations. Continuous wave pump signal with a power of 1 mW at a light wavelength of 1.3 μm propagating along the 1 μm wide SOI waveguide produces a force of about 0.8 pN acting on the cantilever (Fig. 3b). The fully modulated pump signal induces cantilever oscillations with the amplitude

$$X_A = Q |<F>| / 4\pi^2 f_{M0}^2 M_{eff} \qquad (6)$$

In vacuum and low-pressure gases, the amplitude $X_A$ exceeds 7.0 nm (Fig. 3b), which is easily measured thanks to the high linear displacement detection responsivity of the proposed readout scheme.

**Sensitivity**

Sensitivity of the proposed sensor is fundamentally limited by the thermomechanical noise of the cantilever and the noise of the probe optical signal. Thermal force acting on the cantilever has a random phase and a white spectrum with the spectral density $S_F(\omega) = 2M_{eff}k_B\theta\gamma_{eff}/\pi$, where $\theta$ is the temperature of the cantilever, $M_{eff}$ is the effective mass of the cantilever and $\gamma_{eff} = 2\pi f_{M0}/Q$ is the effective damping constant. Spectral density of the corresponding random cantilever displacement is expressed as:

$$S_X(\omega) = \frac{1}{\left(\omega^2 - 4\pi^2 f_{M0}^2\right)^2 + 4\pi^2 f_{M0}^2 \omega^2/Q^2} \frac{S_F(\omega)}{M_{eff}^2} \quad (7)$$

In resonance, $S_X(\omega)$ increases up to $S_{X0} = 5.9 \times 10^{-6}$ nm$^2$/Hz at room temperature, which eventually is converted to the relative intensity noise (RIN) at the photodetector $RIN_X = S_{X0}R_X^2$. When the SiN nanobeam is 50 nm above the silicon waveguide, $RIN_X = -150$ dB/Hz exceeds the inherent RIN of the probe optical signal, which is below $-120$ dB/Hz for low cost semiconductor laser diodes and is in the range between $-180$ and $-150$ dB/Hz for lasers used in telecommunications. At the same time, we should note that the root mean square (r.m.s.) fluctuation amplitude of the cantilever is only 0.97 Å, which is much smaller than the amplitude of the oscillations induced by the 1 mW pump signal if the cantilever is separated from the waveguide by a distance shorter than 150 nm (Fig. 3b). Thus, the all-nanophotonic transduction-actuation scheme does not practically limit the sensitivity of the device, which is mostly determined by mechanical properties of the cantilever and can be evaluated as [31]

$$\delta m = \frac{1}{4\pi R_f X_A} \left[\frac{Qk_B\theta}{3\pi M_{eff}} \left(\frac{\Delta f}{f_{M0}}\right)^3\right]^{\frac{1}{2}} \quad (8)$$

where $\Delta f$ is the measurement bandwidth, which is of the order of the inverse measurement time $1/2\pi\tau < f_{M0}/2Q$ [32]. Accordingly, the considered SiN cantilever placed 50 nm above the silicon waveguide and excited by the fully modulated 1 mW pump signal demonstrates a sensitivity of about $\delta m = 4$ kDa at a measurement bandwidth of 100 Hz, and the sensitivity is improved up to $\delta m = 130$ Da as $\Delta f$ is decreased down to 10 Hz.

**Improvement of the all-nanophotonic gas and biosensor**

The proposed transduction scheme based on a silicon waveguide and a SiN cantilever demonstrates a high linear displacement detection responsivity of more than 0.04 µm$^{-1}$. At the same time, end-to-end coupled nanophotonic waveguide cantilevers can give $R_X$ of about 1 µm$^{-1}$ [19]. The reason for such a high responsivity is that transmission through the gap separating two waveguide cantilevers is very sensitive to the relative position of the waveguide ends thanks to the high mode confinement [19]. In the case of the mechanical cantilever suspended above the waveguide (Fig. 1), the dependence of the transmission coefficient $T$ on the separation distance between the cantilever and the waveguide is the result of the interplay between two opposite effects. The power loss of the guided mode $1 - T$ is proportional to the square of the normalized electric-field amplitude of the guided mode at the position of the cantilever $E_{gy}(h)$ (see equation 2). First, this means that $1 - T$ decreases with the distance $h$ approximately as $\exp(-2h/\zeta_{top})$ and, consequently, $R_X \propto -\exp(-2h/\zeta_{top})/\zeta_{top}$. In other words, the higher mode confinement, which is characterized by $1/\zeta_{top}$, the higher the linear displacement responsivity $R_X$. Second, $R_X$ also depends on the normalized amplitude of the electric field at $x = h$, which can be calculated from the equation $\int_{-\infty}^{+\infty} |E_{gy}(x)|^2 dx = 4\omega/\beta_g c$ [27]. The same interplay between the magnitude of the normalized electric field at $x = h$ and the field gradient is evident in actuation, which is governed by the ponderomotive optical force. If the refractive index contrast between the waveguide core and the claddings is very high, the energy flows mostly in the waveguide core and the amplitude of the electric field in the claddings is much smaller than that in the center of the core, while $1/\zeta_{top}$ can be very large. In the opposite case of the low refractive index contrast, the power flow is spread out over a wide area in the waveguide cross-section and both the normalized field amplitude at the position of the cantilever and the inverse penetration depth into the cladding are very small resulting in a very low responsivity. This demonstrates that strong mode confinement is not a cornerstone for high responsivity and efficient actuation.

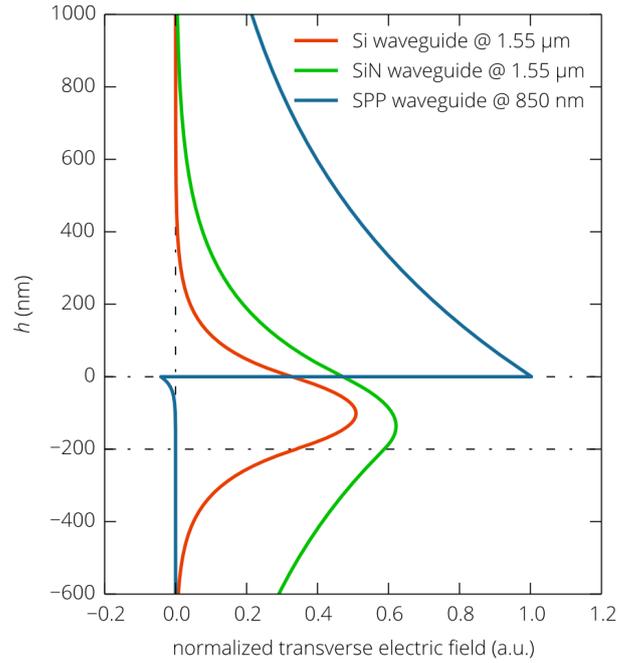

**Figure 4. Normalized field distributions for the fundamental modes of photonic and plasmonic waveguides.** Transverse electric fields for the fundamental TE mode of the Si and SiN photonic waveguides and for the fundamental TM mode of the SPP waveguide. The fields of the modes are normalized to carry the same power. Penetration depth of the optical mode in the SiN waveguide is twice as large as that in the Si one, the intensity of the SPP mode near the metal surface is much higher and the electrical field decays slower than that of the photonics mode of the Si and SiN waveguide.

In contrast to silicon, silicon nitride exhibits much higher chemical stability, while still ensuring compatibility with the CMOS fabrication process. Refractive index of SiN is 1.75 times smaller than that of silicon and the 200 nm thick SiN waveguide provides poorer mode confinement.

Penetration depth of the guided $TE_0$ mode into air is 220 nm compared to 96 nm in the case of the silicon waveguide. Such mode size expansion is accompanied by more uniform field distribution and the normalized electric field at the position of the cantilever is significantly higher than that for the silicon waveguide (Fig. 4). This results in a strong interaction between the nanobeam and the guided mode. At a height of 50 nm, $1 - T$ is 50 times greater than in the case of the silicon waveguide, so is the linear displacement responsivity: $R_X|_{h=50nm} = 0.79$ $\mu m^{-1}$ corresponding to the relative intensity noise at the photodetector $RIN_X = -125$ dB/Hz. This feature favors the use of relatively noisy low cost laser diodes. Also important is that the responsivity decreases quite slowly as the distance between the cantilever and the waveguide increases and it is greater than 0.01 $\mu m^{-1}$ at a height of 650 nm. Very similar trend is observed in the dependence of the actuation force on the cantilever position. The amplitude of oscillations induced by the fully modulated 1 mW pump signal is about 22 nm at $h$ = 50 nm, which is more than one order of magnitude greater than in the case of the end-to-end coupled nanophotonic waveguide cantilevers [19]. The amplitude of induced oscillations slowly decreases with the distance from the metal film, but remains higher than the r.m.s. thermal vibration amplitude until $h$ exceeds 550 nm (Fig. 5b).

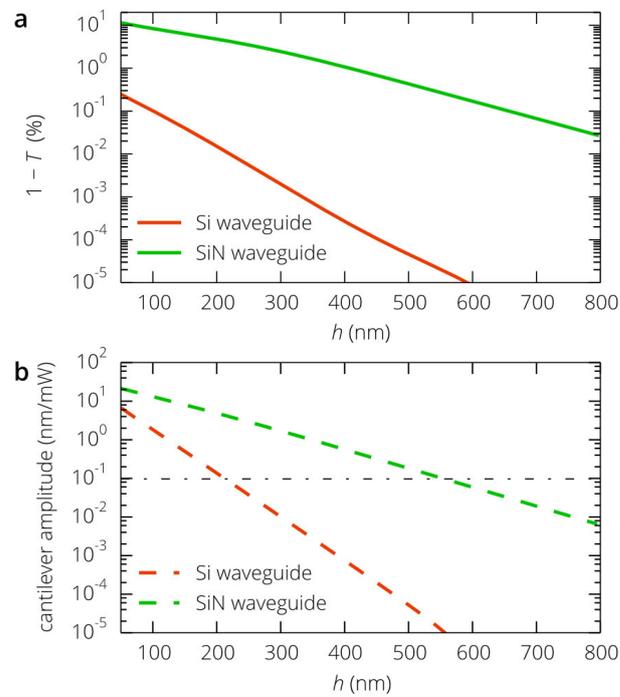

**Figure 5. Properties of the transduction and action scheme based on the SiN waveguide.** (a) Transmittance of the probe optical signal through the waveguide section with the SiN cantilever for the silicon and silicon nitride waveguides at a light wavelength of 1.55 μm. (b) Amplitude of the cantilever mechanical oscillations, induced by the fully modulated pump optical signal at a light wavelength of 1.31 μm, versus the distance to the waveguide surface. The dash-dotted line corresponds to the r.m.s. thermal vibration amplitude (0.097 nm) in the case of the 1 mW pump signal.

The above simulations show that the linear displacement responsivity for photonic waveguides is of the order of 1 $\mu m^{-1}$ and in principle can be further increased by half an order of magnitude. However, insulator and SOI structures suffer from the fact that a significant portion of the electromagnetic field is concentrated in the waveguide core and the power flow in the substrate is higher than in the cover (air or vacuum) layer, since the penetration depth the guided mode with the wavenumber $\beta_g$ into air or vacuum $\zeta_{top} = 1/\sqrt{\beta_g^2 - \omega^2/c^2}$ is several times smaller than that into the substrate $\zeta_{bot} = 1/\sqrt{\beta_g^2 - n_{SiO2}^2 \omega^2/c^2} > \zeta_{top}$. Accordingly, the normalized amplitude of the electric field at the position of the cantilever is fundamentally limited. This problem can be solved only by avoiding field localization in the waveguide core and in the substrate, which is impossible with all-dielectric waveguides. Nevertheless, surface waves can overcome this limitation if the penetration depth of the electromagnetic field into one of the media is much smaller than in the other. At optical frequencies, such a situation is realized for surface plasmon polaritons (SPPs) [33].

SPPs being collective excitations of the conductive electron on the metal surface can be simply thought of as the propagating TM waves with the dispersion relation $\beta_g = \omega/c \times [\varepsilon_m/(1+\varepsilon_m)]^{1/2}$ [33], where $\varepsilon_m$ is the dielectric constant of the metal. Penetration depth of the electromagnetic field into the metal is only about 20 nm and most of the energy typically flows in air. Amplitude of the electric field decreases with the distance from the metal surface as $\exp(-x/\zeta_{top})$, where $\zeta_{top}$ depends strongly on the SPP frequency. For an SPP propagating along the smooth gold [34] or copper [35] surface, $\zeta_{top}$ = 2.65 µm at λ = 1.55 µm, which is greater than the light wavelength in vacuum and, as opposed to the common belief, the SPP is poorly confined to the metal surface. However, $\zeta_{top}$ rapidly decreases as λ decreases and is equal to 705 nm at a light wavelength of 850 nm. It should be noted that in the case of the silicon photonic waveguide we are fundamentally limited in the operating wavelength of light, since silicon strongly absorbs photons with energies greater than the bandgap energy $E_g^{Si}$ = 1.1 eV (λ < 1.1 µm). Plasmonic waveguides, in turn, allow to operate at much shorter wavelengths despite that ohmic losses increase as the confinement to the metal surface increases and the insertion loss rises from 0.74dB/100µm×$L_{WG}$, where $L_{WG}$ is the waveguide length, at λ = 1.55 µm to 3.6dB/100µm×$L_{WG}$ at λ = 850 nm, but still remains at a relatively low level.

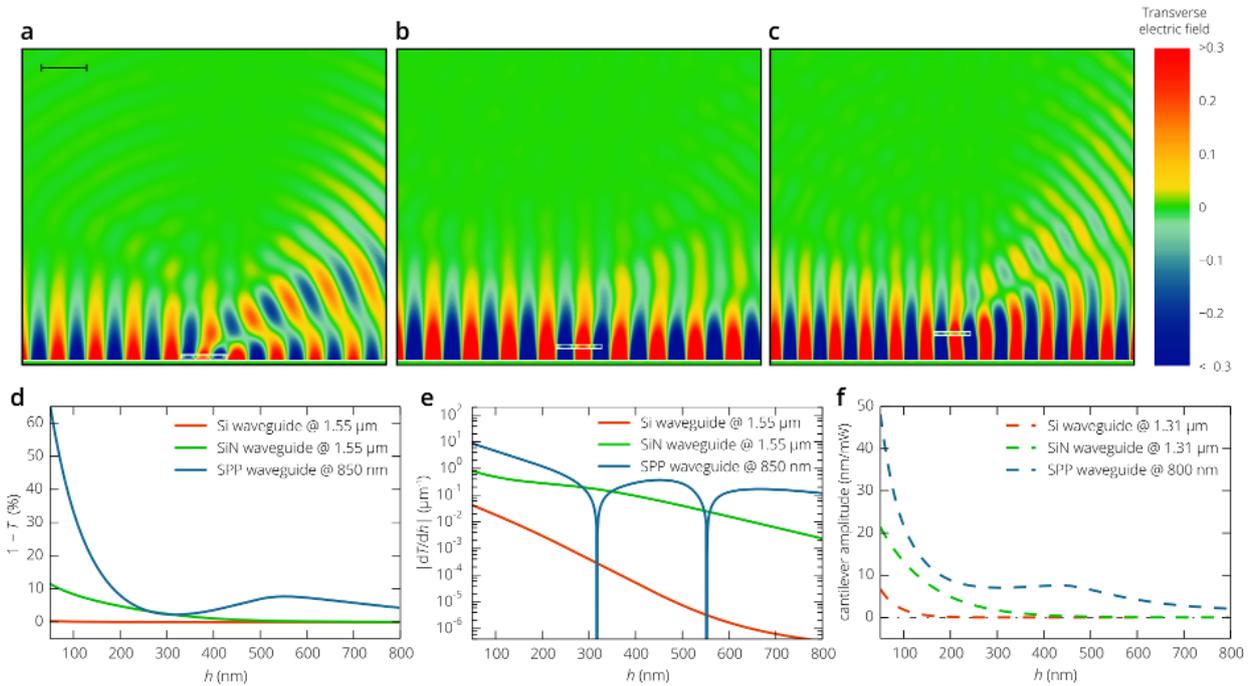

**Figure 6. All-plasmonic biosensor.** Simulated transverse electric field distribution of the SPP passing through the waveguide section with the SiN cantilever at three different positions of the cantilever: h = 40 nm (panel a),

$h = 280$ nm (panel b), $h = 560$ nm (panel c). Panels d and e show the dependences of the transmittance and its derivative d$T$/d$h$ on the distance between the cantilever and the waveguide. As opposed to the Si and SiN photonic waveguides (Fig. 5a), (1-$T$) does not steadily decrease with $h$ increases and well pronounced minima and maxima are observed. These extrema give zero linear displacement responsivity |d$T$/d$h$| at $h = 317$ nm and $h = 552$ nm (see panel e). Panel f presents the amplitude of the cantilever oscillations induced by the fully modulated pump signal per unit power of the pump signal before it is modulated.

Fig. 6a-c presents the simulated transverse electric field of the all-plasmonic gas and biosensor based on the SPP waveguide and SiN cantilever, suspended above the waveguide. In contrast to the Si and SiN photonic waveguides at telecom wavelengths, the penetration depth of the SPP field into air or vacuum ($\zeta_{top} = 705$ nm) is comparable with the light wavelength ($\lambda = 850$ nm) and the profile of the nanobeam provides high overlap between the guided mode and radiation modes of the plasmonic waveguides, which eventually results in the characteristic minima and maxima in the dependence of the transmission coefficient $T$ on the distance from the metal surface (Fig. 6d). At a height of 50 nm, transmission is less than 50%, $1 - T = 0.66$, which is six times greater than in the case of the SiN waveguide and 270 times greater than in the case of the Si waveguide (Fig. 5a). In spite of the fact that the penetration depth of the SPP $\zeta_{top} = 705$ nm is more than seven times larger than that of the TE$_0$ mode of the silicon waveguide ($\zeta_{top} = 96$ nm), $1 - T$ and $|R_X|$ decreases with the distance from the waveguide surface roughly at a similar rate (Fig. 6d-e) due to resonant coupling between guided and radiation modes of the SPP waveguide [28], which gives extrema in Fig. 6d at a distance of 317 nm and 552 nm. In these positions of the cantilever, the linear displacement responsivity is zero and the local maxima are reached at $h = 440$ nm and $h = 680$ nm. But, obviously, the highest responsivity is achieved near the metal surface. Thus, for example, at a height of 50 nm, $|R_X| = 8$ μm$^{-1}$ and corresponds to the relative intensity noise at a photodetector of $-104$ dB/Hz, which is several orders of magnitude higher than the RIN of low cost semiconductor diodes. The responsivity can be easily further improved by operating at shorter optical wavelengths, but this is unavoidably accompanied by increase in the insertion loss of the sensor and the practical limit is dictated by the size and complexity of the whole photonic scheme. As opposed to the responsivity, optical forces are determined only by the field gradient of the guided mode and, at $h = 50$ nm, the amplitude of mechanical oscillations for the cantilever suspended above the plasmonic waveguide is only twice as large as that in the case of the SiN waveguide and seven times greater than in the case of the Si waveguide. However, since the amplitude of electric field decreases with the distance from the waveguide surface slower than for the Si and SiN photonic waveguide (Fig. 4), the amplitude of the cantilever oscillations induced by the fully modulated pump signal with a power of 1 mW surpasses, by more than one order of magnitude, the r.m.s. thermal noise amplitude even at a distance between the nanobeam and the waveguide of greater than 1 μm. At the same time, the responsivity at $h = 1$ μm is equal to 0.06 μm$^{-1}$ and is sufficiently high for practical applications. Thus, the proposed all-plasmonic biosensor demonstrates the highest linear displacement responsivity, the most efficient optical actuation and can operate in a wide range of distances between the nanobeam cantilever and plasmonic waveguide.

**Operation in atmosphere**

In vacuum, the energy of the oscillating cantilever is dissipated due to internal friction and the quality factor of the resonance is of about $Q_{int} = 3500$. This gives the possibility to efficiently excite the cantilever (see equation (6)) and detect changes in mass with single-molecule resolution (see equation (8)). However, for practical applications, operation at atmospheric pressure is desirable, but, as the gas pressure increases, the hydrodynamic drag becomes dominant over the internal dissipation significantly decreasing the quality factor of the mechanical resonator. Firstly, damping is caused by individual gas molecules, which collide with the oscillating cantilever. The corresponding quality factor can be given by [36]

$$Q_B = \frac{3\pi^2 \rho t \upsilon_T f_{M0}}{8p}, \qquad (9)$$

where $p$ is the gas pressure and $\upsilon_T$ is the average thermal velocity of gas molecules. At room temperature in air at a pressure of 1 atm $Q_B = 26.8$, which is more than two orders of magnitude lower than the quality factor limited by internal dissipation effects. Secondly, at high pressure, the density of gas is so large that we cannot neglect interaction between gas molecules and viscous effects must be considered. The hydrodynamic drag force due to viscous friction of the oscillating cantilever with the dense gas per unit length of the cantilever can be written as $F_{visc} = \beta_1 du(y)/dt$, where $u(y)$ is the displacement of the cantilever from the equilibrium position at the distance $y$ from the fixed end and $\beta_1$ is the damping constant, which can be estimated as [37]

$$\beta_1 = 3\pi\mu w + \frac{3}{4}\pi w^2 \sqrt{4\pi\mu\rho_{gas} f_{M0}}. \qquad (10)$$

In this expression, $\mu$ is the dynamic viscosity and $\rho_{gas}$ is the density of the gas. Accordingly, the quality factor associated with the viscous friction is equal to

$$Q_{visc} = 2\pi\rho_{gas} t w^2 f_{M0}/\beta_1. \qquad (11)$$

Finally, since the cantilever is placed above the substrate at a distance of a few hundred nanometers (Fig. 1), an additional drag force appears due to the small gaps between the beam and the waveguide and between the beam and the SiO$_2$ surface. The vibrating beam squeezes the gas film between the beam and substrate and causes the gas to flow towards the beam edges. Following Ref. [37], we can roughly estimate the squeeze-film quality factor as

$$Q_{sq} = 2\pi f_{M0} \rho t h^3 / \mu w^2 \qquad (12)$$

and calculate the total quality factor as $Q = 1/(1/Q_{int} + 1/Q_B + 1/Q_{visc} + 1/Q_{sq})$. Fig. 7a shows that the damping force becomes very strong, when the cantilever is separated from the substrate by a distance of a few nanometers and $Q$ becomes much smaller than unity. As the cantilever is moved away from the waveguide, the quality factor steadily increases and, at a distance of 1 μm, the gap between the waveguide and the SiN beam does not have any impact on cantilever damping and $Q \approx 1/(1/Q_B + 1/Q_{visc}) = 15$.

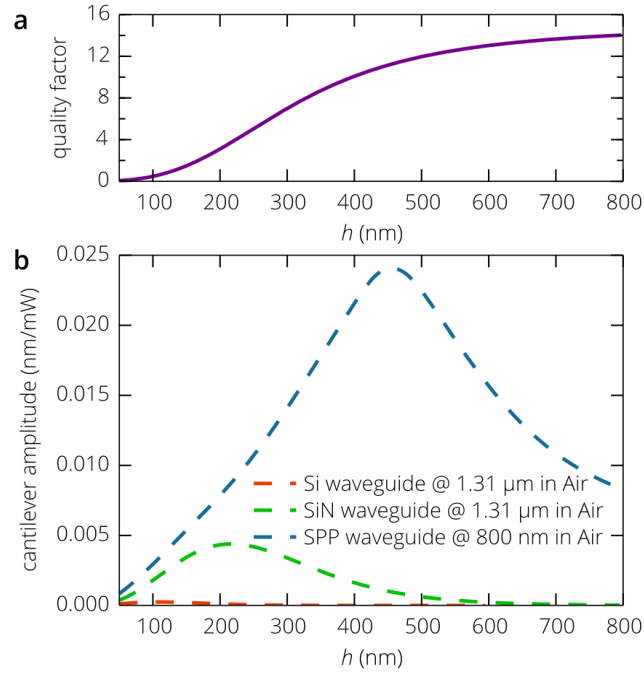

**Figure 7. Operation of the all-nanophotonic biosensor in atmosphere.** (a) Dependence of the quality factor of the mechanical resonator on the cantilever position above the nanophotonic waveguide in atmosphere. (b) Amplitude of mechanical oscillations induced by the fully modulated pump optical signal as a function of the separation distance between the SiN cantilever in air and the waveguide surface.

In contrast to the case of induced oscillations in vacuum (Fig. 6f), the vibration amplitude, being inversely proportional to the quality factor, does not monotonically decrease as $h$ increases (Fig. 7b) and for all three waveguides the optimal positions of the SiN beam above the waveguide are clearly observed. Silicon waveguide provides the best confinement of the optical mode to the waveguide core and the highest gradient of the electromagnetic field becomes practically useless at atmospheric pressure: the maximum amplitude is achieved at $h = 110$ nm and is equal to only $2 \times 10^{-4}$ nm per milliwatt of the pump optical signal, while $R_X$ is less than 0.015 μm$^{-1}$ and $Q = 0.6$. SiN waveguide demonstrates much better characteristics: the amplitude exceeds $4 \times 10^{-3}$ nm/mW for $h$ in the range from 200 nm to 300 nm, which combined with the high linear displacement responsivity gives a sensitivity per power of the pump signal of (2–4)$\times 10^5$ Da·mW at a measurement bandwidth of 100 Hz . In order to improve the sensitivity, one needs to operate with higher mechanical quality factors. Due to the squeeze-film damping, this can be possible only by moving the cantilever away from the waveguide, but as $h$ increases, both the displacement responsivity and the oscillation amplitude rapidly decrease and the best sensitivity for the considered SiN waveguide is achieved at a distance of 140 nm from the waveguide surface, where the quality factor is about unity. However, this mass sensitivity is associated with the inherent properties of the waveguide and the cantilever in air and the RIN at the photodetector will be limited by the laser, as follows from Eq. (8), unless the laser noise is of about −180 dB/Hz.

As opposed to photonic modes of dielectric waveguides, the linear displacement responsivity and the optical force acting on the cantilever are not monotonically decreasing functions of the gap distance between the waveguide and the cantilever (Fig. 6e-f). They exhibit maxima at $h \approx 440$ nm giving a possibility to operate with a mechanical quality factor of about 11. In addition, relatively high penetration depth of the electromagnetic field into air and high field confinement in the air region above the metal surface (for details see section *Improvement of the all-nanophotonic gas and biosensor*) result in a high magnitude of these maxima, so that the maximum sensitivity of the biosensor per power of the pump signal is equal to $6.5 \times 10^4$ Da·mW at a measurement bandwidth of 100 Hz. In addition, this is achieved at the moderate mechanical quality factor diminishing the influence of the laser noise on the sensitivity of the proposed biosensor and giving a possibility to use freely lasers with the RIN of up to −145 dB/Hz without reduction in accuracy of mass detection.

## Discussion

We have proposed a novel scheme of highly integrated NEMS biosensor with all-nanophotonic transduction and actuation. Such a configuration consists of the photonic waveguide and the nanobeam cantilever placed above it at distance of a few hundreds of nanometers (Fig. 1). Mechanical vibrations of the cantilever are excited by the sinusoidally modulated pump optical signal propagating along the waveguide and the same waveguide serves as an optical transducer: the vibrating cantilever controls the transmission of the continuous wave probe optical signal through the section with the suspended nanobeam. In addition, optical transduction does not use interferometric effects and therefore low noise incoherent light sources with relatively broad emission spectrum can be utilized. At the same time, even in the case of the single cantilever with a beam cross-section of only 0.09 μm$^2$, the linear displacement responsivity of the system exceeds 0.8 μm$^{-1}$ for the SiN waveguide and 9 μm$^{-1}$ for the SPP one. Thanks to such a high responsivity and efficient optical actuation, mass-detection sensitivity is limited by the thermomechanical noise of the cantilever rather than the properties of the optical source. For the cantilever suspended at a distance of 50 nm above the SiN waveguide in vacuum, $\delta m$ is of about 6.5 kDa at a measurement bandwidth of $\Delta f$ = 100 Hz and a pump signal power of $P$ = 200 μW, while for the SPP waveguide under the same conditions $\delta m \approx 2.9$ kDa. In atmosphere, NEMS are difficult to operate due the very low quality factor caused by the air flow and squeeze-film damping. However, we have shown that conventional low noise lasers do not limit the sensitivity of the scheme based on the SPP waveguide and $\delta m$ is about 65 kDa at $\Delta f$ = 100 Hz and $P$ = 1 mW, which is small enough to detect a single biomolecule. Thus, being easily manufactured, extremely compact and compatible with standard CMOS fabrication processes, the proposed biosensor scheme forms the backbone for parallel monitoring and analysis with a single-molecule resolution at the chip scale.

**Acknowledgements**

This work is supported by the Ministry of Education and Science of the Russian Federation (no. 16.19.2014/K and CCU MIPT RFMEFI59414X0009).

## Author contributions

D.Y.F conceived the idea and wrote the manuscript. Y.V.S. performed the numerical simulations and contributed to the manuscript preparation. Both authors did theoretical calculations and analyzed the results.

## Additional Information

**Competing Financial Interests statement**

The authors declare no competing financial interests.